\documentclass[12pt]{article}
\usepackage{graphicx,epsfig}
\usepackage{color}
\usepackage{cite}
\title{Formation and condensation of excitonic bound states in the
generalized Falicov-Kimball model}
\author{Pavol Farka\v sovsk\'y\\
Institute  of  Experimental  Physics,  Slovak   Academy   of
Sciences\\
Watsonova 47, 043 53 Ko\v {s}ice, Slovakia}
\date{}
\begin{document}
\baselineskip=24pt
\maketitle

\begin{abstract}
The density-matrix-renormalization-group (DMRG) method  and the Hartree-Fock (HF) 
approximation with the charge-density-wave (CDW) instability are used to study 
a formation and condensation of excitonic bound states in the generalized 
Falicov-Kimball model. In particular, we examine effects of various factors, 
like the $f$-electron hopping, the local and nonlocal hybridization, as well as 
the increasing dimension of the system on the excitonic momentum 
distribution $N(q)$ and especially on the number of zero momentum 
excitons $N_0=N(q=0)$ in the condensate. It is found that the negative 
values of the $f$-electron hopping integrals $t_f$ support the formation 
of zero-momentum condensate, while the positive values of $t_f$ have the fully 
opposite effect. The opposite effects on the formation of condensate exhibit 
also the local and nonlocal hybridization. The first one strongly supports  
the formation of condensate, while the second one destroys it completely. 
Moreover, it was shown that the  zero-momentum  condensate remains robust 
with increasing dimension of the system.  

\end{abstract}

\newpage
\section{Introduction}
The formation and condensation of excitonic bound states of conduction-band
electrons and valence-band holes  belongs surely to  one of the most
exciting ideas of contemporary solid state physics. Although the excitonic
condensation has been predicted a long time ago~\cite{theo}, no conclusive 
experimental proof of its existence has been achieved yet. The latest 
experimental studies of materials with strong electronic correlations 
showed however, that there are a few promising candidates for the experimental 
verification of the excitonic condensation. The first one is the mixed 
valence compound $TmSe_{0.45}Te_{0.55}$, where detailed studies of 
the pressure-induced semiconductor-semimetal transition indicate that 
excitons are created in a large number and condense below 20~K~\cite{Wachter}.
Moreover, the charge-density-wave (CDW) state observed in the layered 
transition-metal dichalcogenide $1T-TiSe_2$ was claimed to be of 
excitonic origin~\cite{Monney}. Quite recently, as a further candidate 
for the excitonic state, a quasi one-dimensional system $Ta_2NiSe_5$
has raised and attracted much experimental attention~\cite{Wakisaka}.
In this material the flat band was observed by the ARPES 
experiment, which was interpreted to be due to excitonic 
condensation. These results have stimulated further experimental 
and theoretical studies with regard to the formation and possible
condensation of excitonic bound states of electron and holes
in correlated systems. At present, it is generally accepted 
that the minimal theoretical model for a description of excitonic 
correlations in these material could be the Falicov-Kimball
model~\cite{Falicov} and its 
extensions~\cite{B1,B2,Z1,Phan,Seki,Z2,Kaneko1,Kaneko2,Ejima,Fark1}. 
In its original form, the Falicov-Kimball model 
describes a two-band system of the itinerant $d$ electrons 
(with the nearest-neighbor $d$-electron hopping constant
$t_d$) and the localized $f$ electrons that interact only via a local 
$f$-$d$ Coulomb interaction $U$:
\begin{equation}
H_0=-t_d\sum_{\langle i,j \rangle}d^+_id_j+U\sum_if^+_if_id^+_id_i+E_f\sum_if^+_if_i,
\end{equation}
where $\alpha^+_i$ and $\alpha_i$ are the creation and annihilation
operators of spinless electrons in the $\alpha=\{d,f\}$ orbital 
at site $i$ and $E_f$ is the position of the $f$-level energy. 
In what follows we consider $t_d=1$ and all energies 
are measured in units of $t_d$.  

Since the local $f$-electron  number $f^+_if_i$ is strictly
conserved quantity, the $d$-$f$ electron coherence cannot be
established in this model. This shortcoming can be overcome 
by including an explicit local hybridization 
$H_V=V\sum_id^+_if_i+f^+_id_i$ between the $d$
and $f$ orbitals.  However, the hybridization between $d$ and $f$
orbitals is not the only way to develop $d$-$f$ coherence. 
Analytical and numerical studies of Batista et al.~\cite{B1,B2} showed 
that a finite $f$-electron bandwidth $H_{t_f}=-t_f\sum_{<i,j>}f^+_if_j$
also induces $d$-$f$ coherence, and thus it can lead to an excitonic 
condensate even in the absence of direct $d$-$f$ hybridization.  
Later this model  has been used extensively to describe 
different phases in the ground state and specially properties of 
the excitonic phase~\cite{Z1,Phan,Seki,Z2,Kaneko1,Kaneko2,Ejima}. 
It was found that the ground state phase diagram    
exhibits very simple structure consisting of only four phases, and
namely, the full $d$ and $f$ band insulator, the excitonic   
insulator, the CDW and the staggered orbital order. 
The excitonic insulator is characterized by a nonvanishing    
$\langle d^+f \rangle$ average. The CDW is described by a periodic
modulation in the total electron density of both $f$ and $d$ electrons,
and the staggered orbital order is characterized by a periodic modulation 
in the difference between the $f$ and $d$ electron densities.

An extension of this model with local hybridization between the 
$f$ and $d$ orbitals has been studied very recently in our work~\cite{Fark1}.
The numerical analysis of the excitonic momentum distribution 
$N(q)=\langle b^+_qb_q\rangle$ (with $b^+_q=(1/\sqrt{L})\sum_k
d^+_{k+q}f_k$, where L denotes the number of lattice sites)
showed that this quantity diverges for $q=0$, signalizing a
Bose-Einstein condensation of preformed excitons.
Moreover,  it was found that the density of zero momentum excitons
$n_0=\frac{1}{L}N(q=0)$ (as well as the total exciton density
$n_T=\frac{1}{L}\sum_q N(q)$) depends strongly on the values of 
the Coulomb interaction $U$ and that already for relatively small values of 
$U$ ($U \sim 4$) the significant fraction of $n_0/n_T\sim0.5$ excitons is
in the zero-momentum state. 

In the current paper we examine in detail how robust is the excitonic state 
against the changes of other model parameters. In particular, we consider
effects of the non-zero f-electron hopping $H_{t_f}$, the non-local 
hybridization $H_{non}=\sum_{<i,j>}V_{i,j}d^+_if_j+H.c.$ 
and the increasing dimension of the system, which together, but also 
separately model more realistically the physical situation in real 
rare-earth compounds than the model considered in  our previous paper 
($t_f=0, E_f=0, D=1$, $H_{non}=0$).
The crucial role of these factors in a correct description 
of ground state properties of these materials has been already confirmed 
by previous analytical and numerical studies. Indeed, it was found, that 
the  f-band (f-electron hopping)  of the opposite parity than the d-band, 
stabilizes the excitonic phase, but only in the dimension $D>1$~\cite{B1,B2}, 
while the nonlocal hybridization (in the mean-field 
approximation) fully destroys the excitonic phase~\cite{Brydon}. 
It should be noted that the comfirmation (refutation) of the last 
result by exact calculations is of crucial importance for 
a correct description of ground states properties of some 
rare earth systems and especially the mixed valence compounds,
like $SmB_6$~\cite{Batko}. In these materials the 
local hybridization is forbidden and only the nonlocal one 
(with inversion symmetry) between the  nearest-neighbour 
f and d orbitals, is allowed~\cite{Czycholl}. 
On the base of the above mentioned facts we expect the strong 
effects of these factors also on the formation and condensation
of the excitonic bound states. Besides these factors, we examine
also the influence of the f-level position on the density
of $d-f$ excitons, since the simple parametrization 
between the f-level position and external pressure
$E_f\sim p$, widely used in the literature~\cite{Gon}  allows us 
to predict, at least qualitatively, the behaviour of this
quantity with the external pressure.

\section{Results and discussion}
\subsection{DMRG results}

To examine effects of the above mentioned factors on the formation 
and condensation of excitonic bound states in the generalized 
Falicov-Kimball model we have performed exhaustive DMRG studies 
of the model Hamiltonian $H=H_0+H_{t_f}+H_V+H_{non}$ for a wide range 
of the model parameters at the total electron density 
$n=n_d+n_f=1$. In all examined cases  we typically keep up  
to 500 states per block, although in the numerically more difficult cases,
where the DMRG results converge slower, we keep up to 1000 states. 
Truncation errors~\cite{White,Malvezzi}, given by the sum of the density matrix
eigenvalues of the discarded states, vary from $10^{-6}$ in the worse 
cases to zero in the best cases.

Let us start a discussion of our numerical results for the case of
$H_{non}=0$. The DMRG results for this case are summarized in Figs.~1-4. 
\begin{figure}[h!]
\begin{center}
\includegraphics[width=7.5cm]{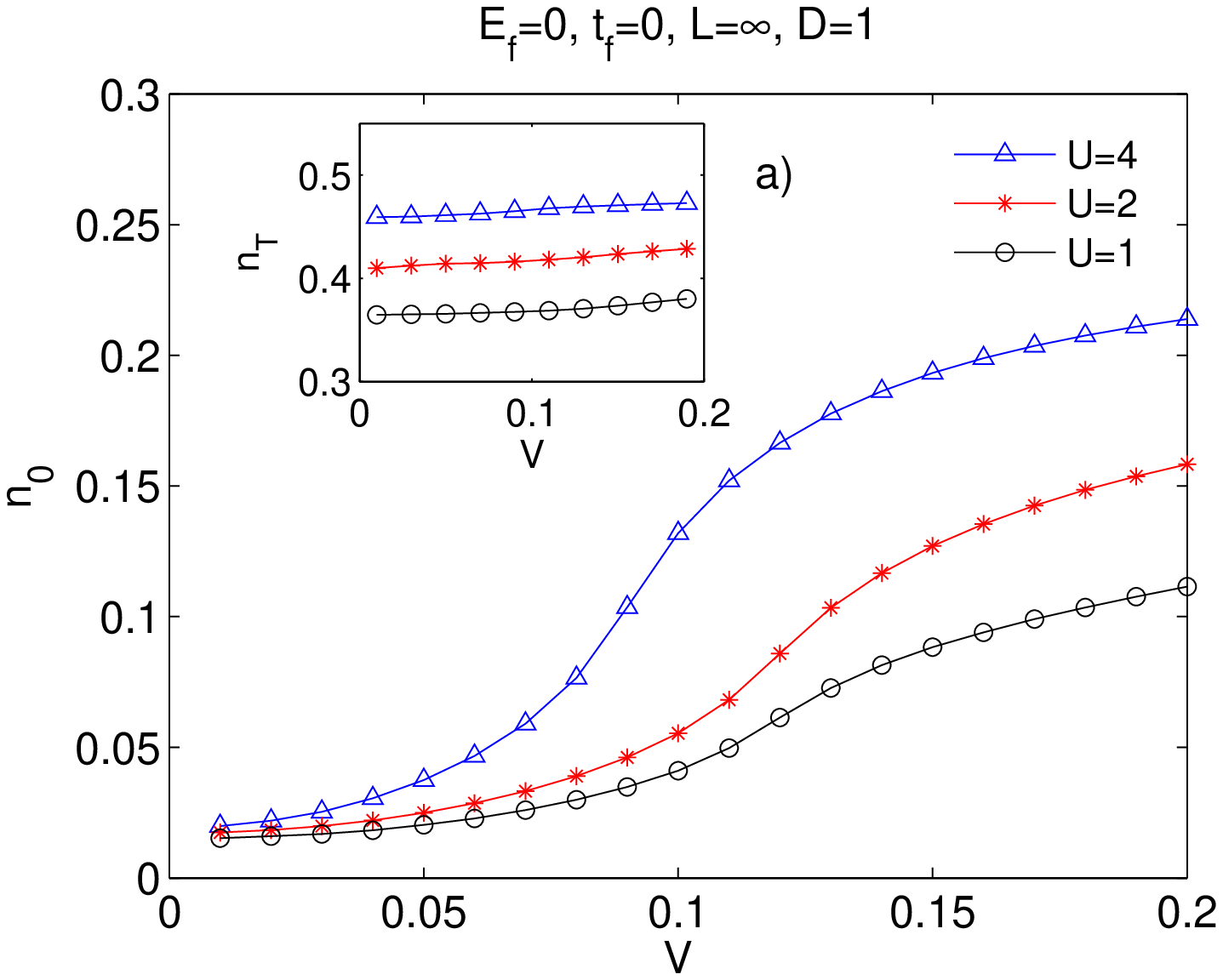}
\includegraphics[width=7.5cm]{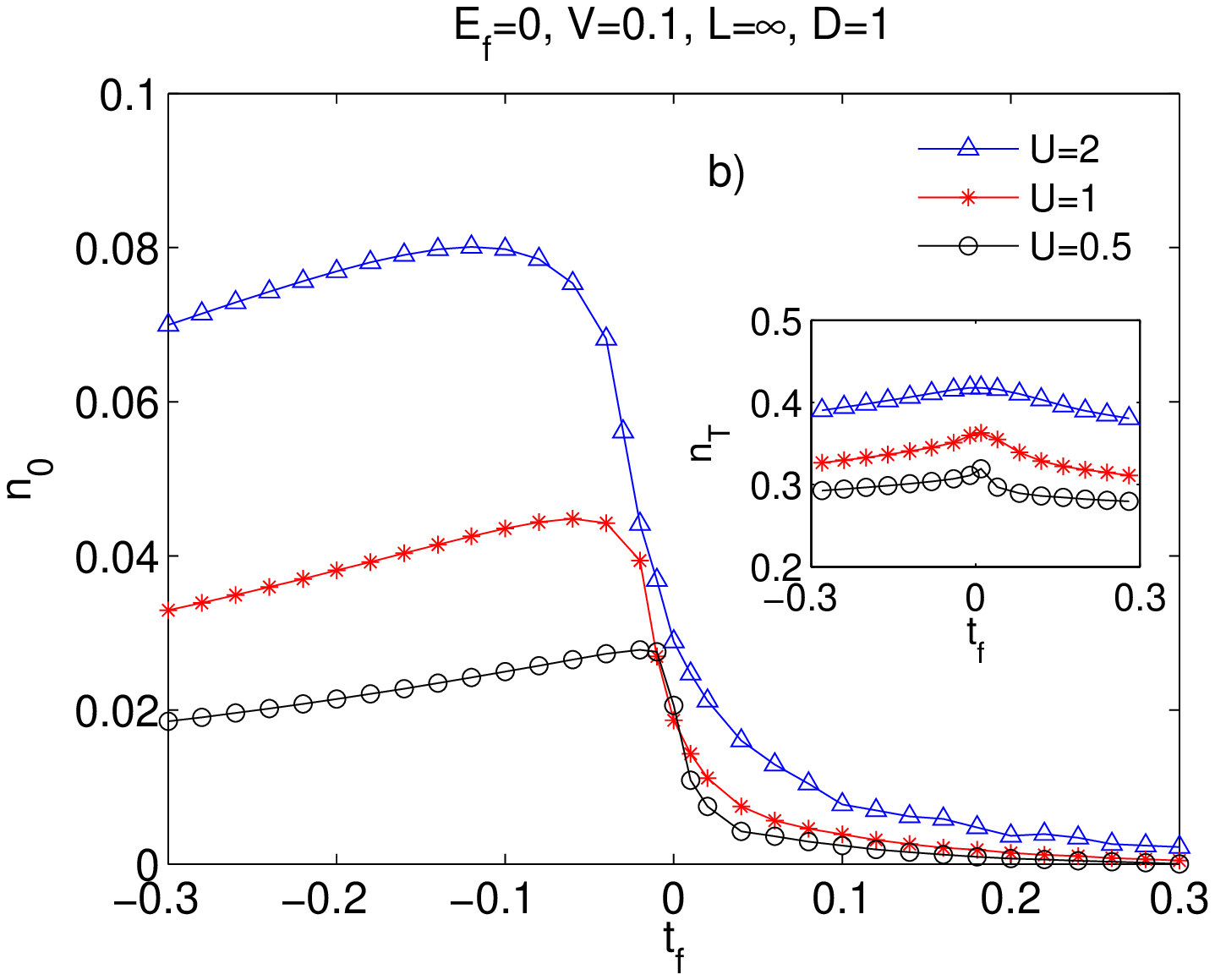}
\end{center}
\vspace*{-0.8cm}
\caption{\small a) The density of zero-momentum excitons $n_0$ 
and the total exciton density $n_T$ as functions of
local hybridization $V$ calculated by DMRG method for three different values of 
the interband Coulomb interaction $U$ ($E_f=0,t_f=0,L=\infty,D=1$).
b) $n_0$  and $n_T$ as functions of $t_f$ calculated for three different values of $U$
($E_f=0,V=0.1,L=\infty,D=1$).}
\label{fig1}
\end{figure}
Fig.~1a shows the bahaviour of the density of zero momentum excitons
$n_0$ as a function of local hybridization $V$ for several values of 
the interband Coulomb interaction $U$. One can see that both, the local
interband hybridization $V$ as well as the local Coulomb interaction $U$
strongly support the condensation of preformed excitons to the zero-momentum
state. In all examined cases the density of zero momentum excitons
$n_0$ is a monotonically increasing function of $V$ which character
changes obviously at some critical value of $V=V_c$ 
($n_0 \sim V^{\nu}$, with $\nu>1$ for $V<V_c$ and
$n_0 \sim V^{\nu}$, with $\nu<1$ for $V>V_c$).
Comparing these results with ones obtained in our previous paper for 
the phase boundary between the CDW and homogeneous phase~\cite{Fark1}, 
the origin of this different behaviour is obvious. While below $V_c$ the 
excitonic phase coexists with the $CDW$ phase, above $V_c$ it coexists
with the homogeneous one, what leads to a different power law  
behaviour of $n_0$ for $V<V_c$ and $V>V_c$. Unlike $n_0$, 
the total exciton density $n_T$ depends only very weakly on 
$V$ (see the inset in Fig.~1a) indicating that the formation 
of excitons is primarily driven by the interband Coulomb interaction $U$.

With regard to the situation in real materials, where always exists
a finite overlap of $f$ orbitals on neighbouring sites, it is more
interesting to ask what happens in the case when a finite $f$ bandwidth
will be considered. 
In accordance with some previous theoretical studies, which 
documented strong effects of the parity of $f$ band on the stability 
of the excitonic phase~\cite{B1,B2} we have examined the model 
for both, the positive (the even parity) and negative 
(the odd parity) values of the $f$-electron hopping integrals $t_f$.
The results of our non-zero $t_f$ DMRG calculations for $n_0$ are displayed 
in Fig.~1b and they  clearly demonstrate that the zero-momentum 
condensate is suppressed in the limit of positive values of $t_f$,
while it remains robust for negative values of $t_f$.
This result is intuitively expected since our previous 
Hartree-Fock (HF) results~\cite{Fark2} showed that only 
the negative values of $t_f$ stabilize the ferroelectric phase, 
while the positive ones stabilize the antiferroelectric phase.
The effect of $t_f$ is especially strong for $U$ small, where continuous but 
very steep changes of $n_0$ are observed for $t_f \to 0^+$.
Contrary to this, the total exciton density $n_T$
exhibits only a weak dependence on the $f$-electron hopping 
parameter $t_f$, over the whole interval of $t_f$ values.

Till now we have presented results exclusively for $E_f=0$.
Let as now discuss briefly the effect of change of the $f$-level
position. This study is interesting also from this point of 
view that taking into account the parametrization between the
external pressure and the position of the $f$ level ($E_f \sim p $),
one can deduce from the $E_f$ dependences of the ground state 
characteristics also their $p$ dependences, at least qualitatively~\cite{Gon}.
The resultant $E_f$ dependences of the density of zero momentum 
excitons $n_0$ obtained by DMRG method are shown in Fig.~2a
for several values of $V$ and $U=0.5$. 
\begin{figure}[h!]
\begin{center}
\includegraphics[width=7.5cm]{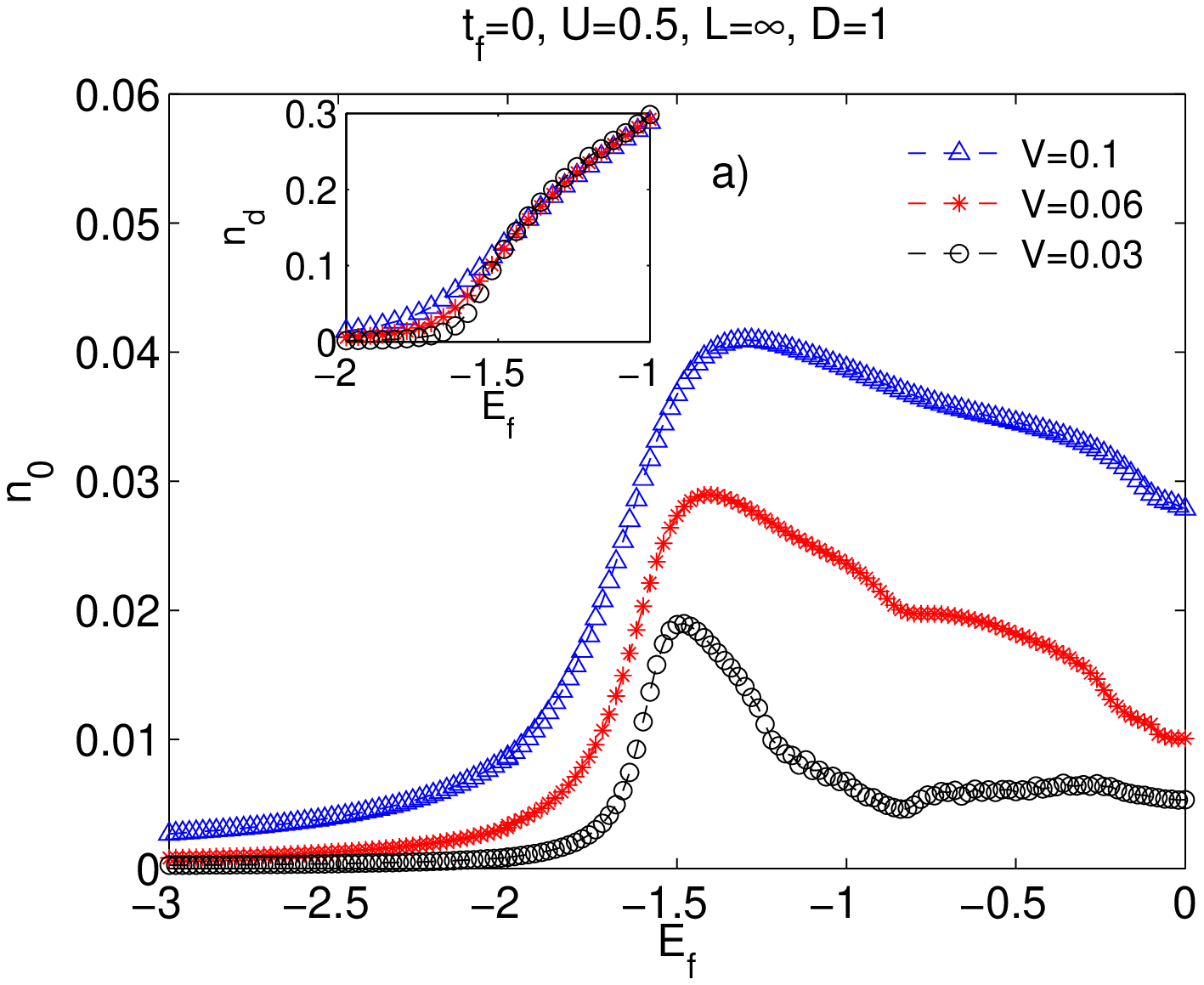}
\includegraphics[width=7.5cm]{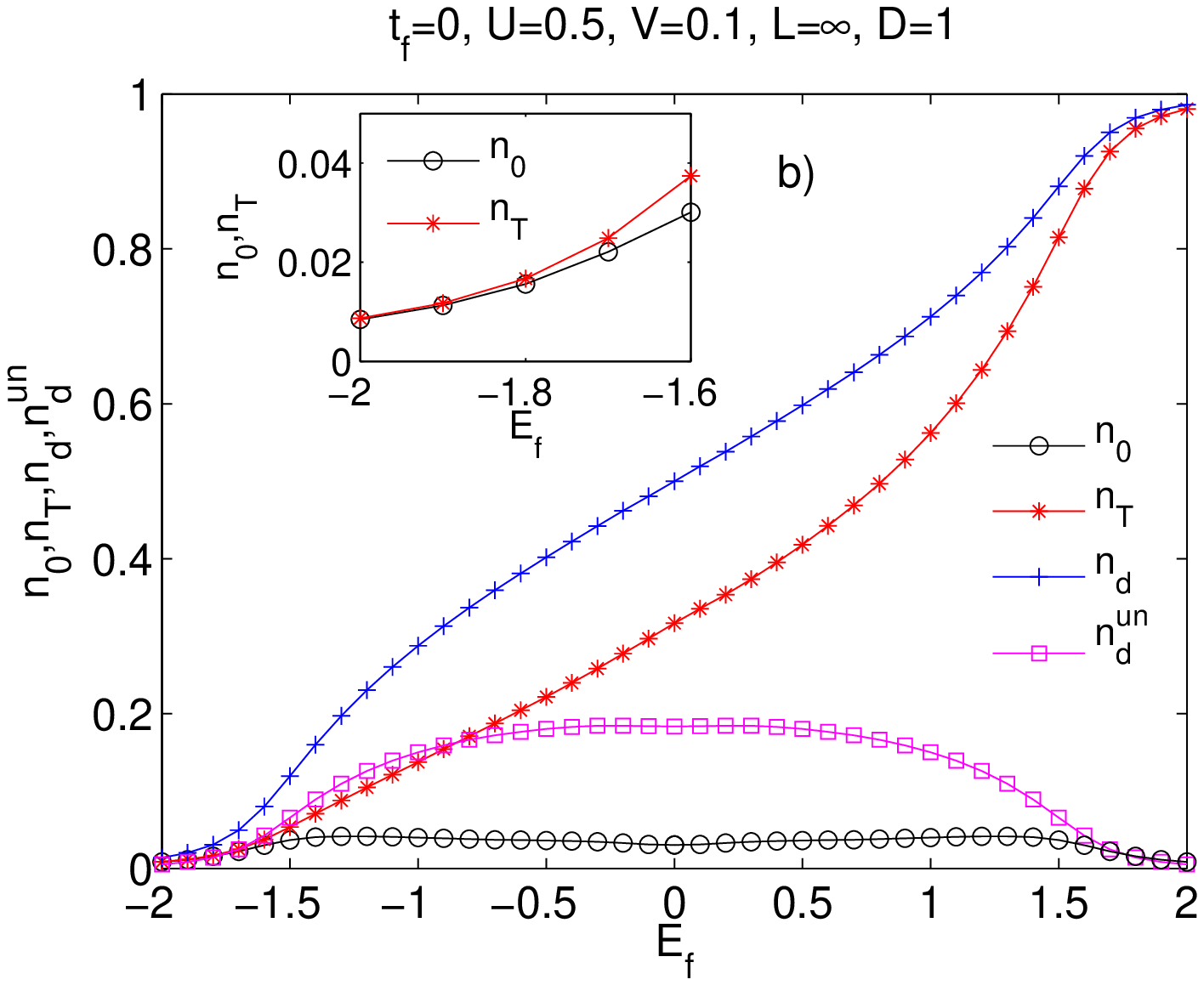}
\end{center}
\vspace*{-0.8cm}
\caption{\small 
a) $n_0$ as a function of $E_f$ calculated by DMRG method for three different values of
$V$ ($t_f=0, U=0.5, L=\infty, D=1$). The inset shows the density of $d$ electrons 
$n_d$  near $E_f=-1.5$.
b) $n_0, n_T, n_d$ and $n^{un}_d=n_d-n_T$ as functions of  $E_f$
calculated for $t_f=0, U=0.5, V=0.1, D=1$ and $L=\infty$.
The inset shows the behaviour of $n_0$ and $n_T$ near $E_f=-2$.}
\label{fig2}
\end{figure}
One can see that the density
of zero momentum excitons is nonzero over  the whole interval
of $E_f$ values. Moreover, we have found that the values of $n_0$ 
are extremely enhanced in the region near $E_f\sim -1.5$,
what is obviously due to the significant enhancement of the 
$d$ electron population in the $d$ band (see the inset in Fig.~2a).

To describe, in more detail, the process of formation
of excitonic bound states with increasing $E_f$, we have
plotted in Fig.~2b, besides the density od zero momentum 
excitons $n_0$, also the total exciton density $n_T$,
the total $d$-electron density $n_d$
and the total density of unbond $d$ electrons $n^{un}_d=n_d-n_T$.
It is seen (see the inset in Fig.~2b) that below $E_f \sim -1.8$,
$n_0$ and $n_T$ coincides what means that the excitonic insulator 
in this region is practically completely driven by the condensation
of zero-momentum excitons. Above this value $n_T$ starts to 
increase sharply, while $n_0$ tends to its maximum at $E_f \sim -1.3$
and then gradualy decreases to its minimum at $E_f=0$.
Similar behaviour with increasing $E_f$ exhibits also 
the density of unbond $d$ electrons $n^{un}_d$,
however the values of $n^{un}_d$ are several times 
larger than $n_0$. It is interesting to note that although 
the total exciton density $n_T$ inreases over the whole 
interval of $E_f$ values, the number of unbond
$d$ electrns remains practically uchanged over the wide 
range of $E_f$ values (from $E_f=-1$ to $E_f=1$), since 
its decrease, due to the formation of excitonic pairs,
is compensated by the increase of $n_d(E_f)$.  
Thus we can conclude that in the pressure induced case,
when the $f$-level energy shifts up with applied 
pressure~\cite{Gon}, the model is able to describe,
at least qualitatively, the increase in the total density of 
excitons with external pressure and the increase or decrease
(according to the initial position of $E_f$ at ambient pressure)
in the $n_0$ and $n^{un}_d$.

As already mentioned, physically the most interesting 
case corresponds however to $H_{non} > 0$. The importance of this 
term emphasizes the fact that the on-site hybridization $V$
is actually forbidden in real $d$-$f$ systems for parity
reasons. Istead the on-site hybridization, one has to consider
in these materials the non-local hybridization with 
inversion symmetry $V_{i,j}=V_{non}(\delta_{j,i-1}-\delta_{j,i+1})$
that leads to $k$-dependent hybridization of the opposite parity
than corresponds to the $d$ band ($V_k\sim sin(k)$)~\cite{Czycholl}.
A straightforward extension of the one-dimensional 
results to two dimensions yields 
$V_{i,j}=V_{non}[\delta_{i_x,j_x}(\delta_{i_y,j_y+1}-\delta_{i_y,j_y-1})+
                 \delta_{i_y,j_y}(\delta_{i_x,j_x+1}-\delta_{i_x,j_x-1})]$,
where any site on the lattice is given by 
${\bf R}_i=i_xa{\bf \hat{x}}+i_ya{\bf \hat{y}}$ and $a$ is the 
lattice constant.
 
Typical examples of $1/L$ dependence of the excitonic momentum 
distribution $N(q=0)$ obtained for three representative values of 
the interband Coulomb interaction and two values of $f$-electron
hopping in the one dimension are displayed in Fig.~3a and Fig.~3b.
\begin{figure}[h!]
\begin{center}
\includegraphics[width=7.5cm]{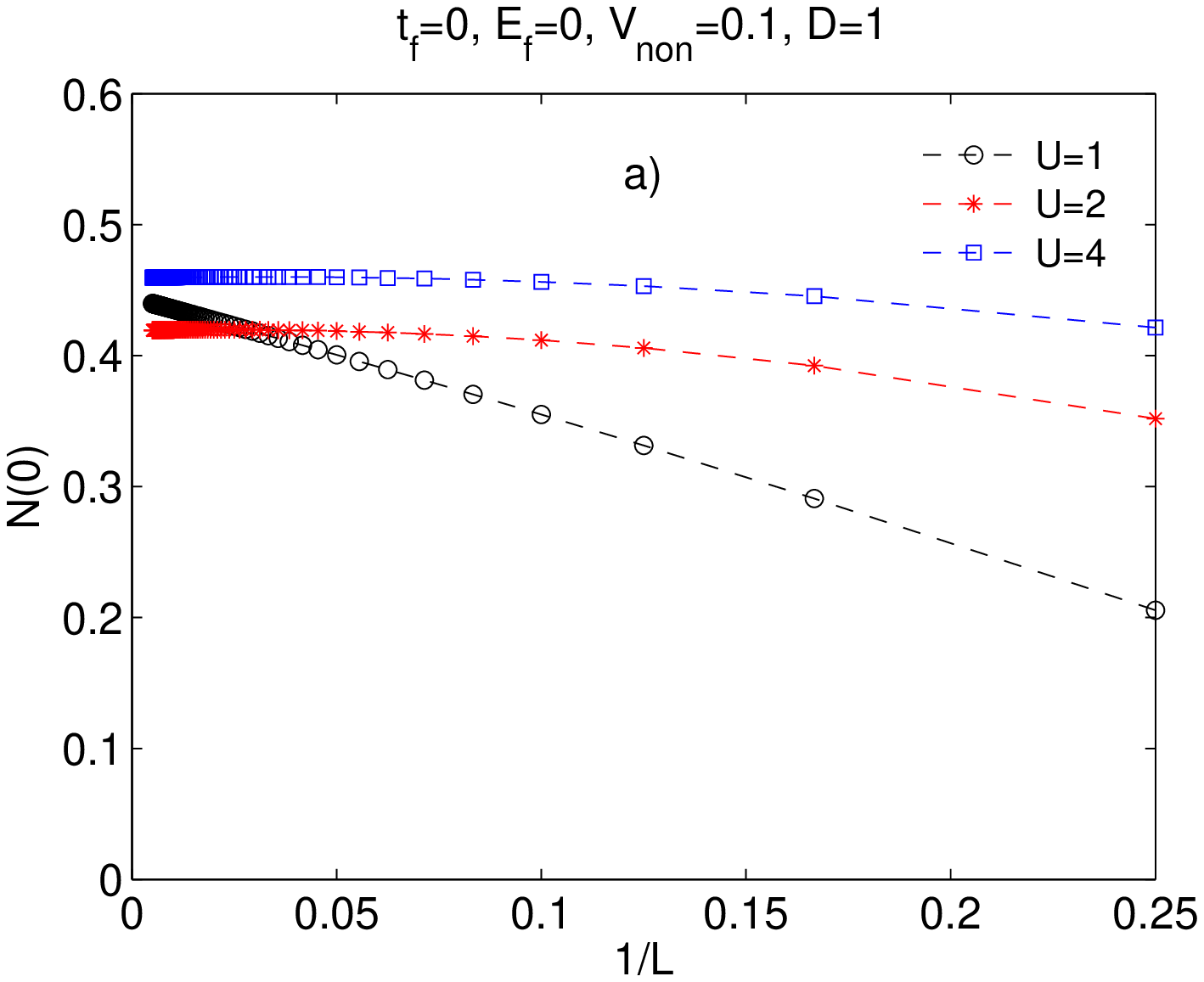}
\includegraphics[width=7.5cm]{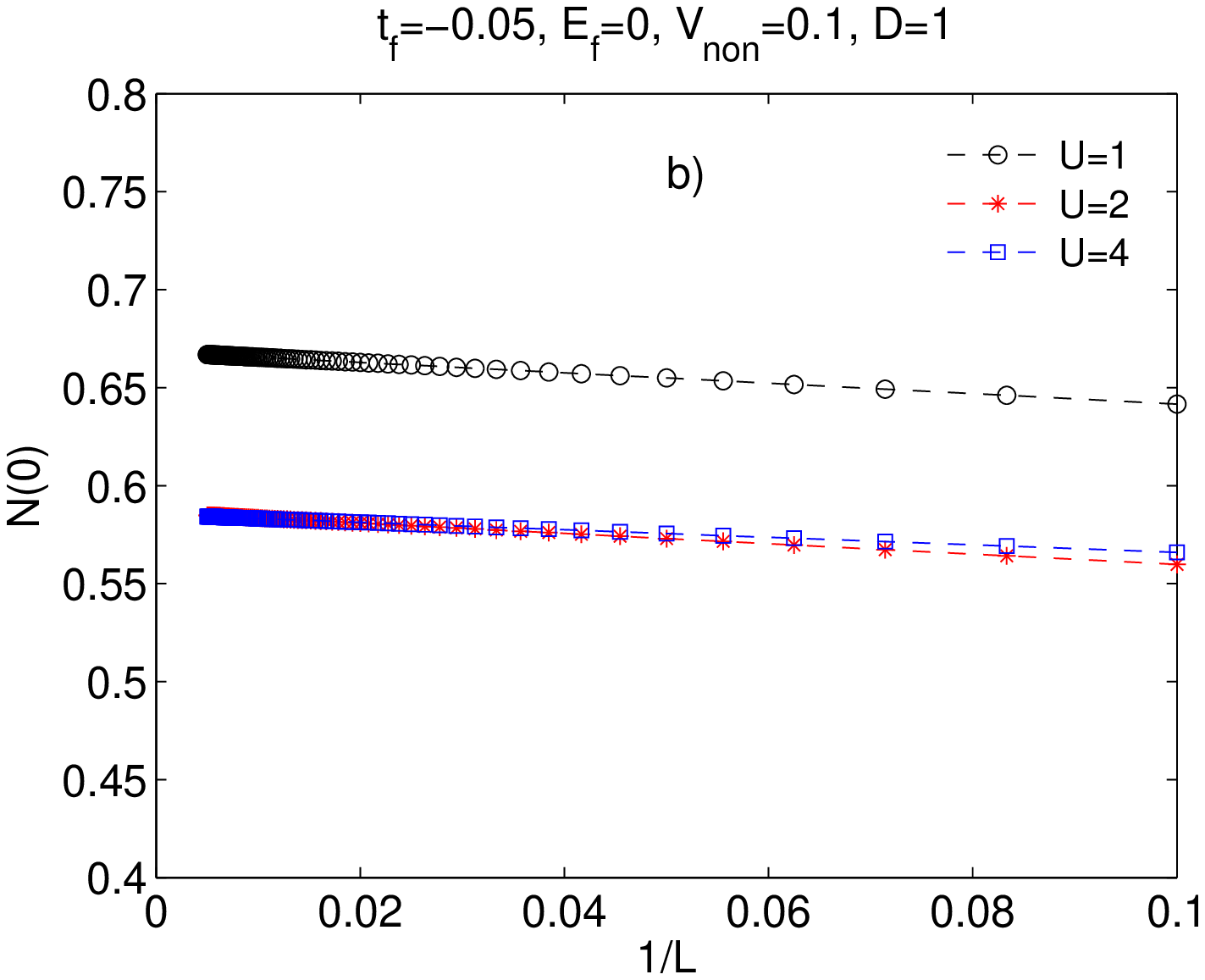}
\end{center}
\vspace*{-0.8cm}
\caption{\small 
$N(0)$ as a function of $1/L$ calculated by DMRG method for three different values of
$U$ and two different values of $t_f$ ($E_f=0, V_{non}=0.1$).
}
\label{fig3}
\end{figure}
These results clearly demonstrate   
that there is no sign of divergence in the $1/L$-dependence of 
$N(0)$  neither for $t_f=0$ nor for $t_f=-0.05$ and thus there 
is no signal of forming the Bose-Einstein condensate in the presence 
of non-local hybridization with the inversion symmetry.
On the base of these results we can conclude that possible 
candidates for the appearance of the Bose-Einstein condensation 
of excitons in real $d$-$f$ materials, are only systems with 
local hybridization that supports the formation of the 
Bose-Einstein condensate, but not systems with
non-local hybridization (of the inversion symmetry), what
strongly limits the class of materials, where this phenomenon
could be observed. On the other hand, it should be noted that although 
the local hybridization is forbidden in real $d$-$f$ electron systems, 
there is still a possibility to induce a finite local hybridization 
by some additional, not fully electronic mechanisms. Such an additional
mechanism could be, for example, the electron-phonon interaction
$H_{el-ph}$. As shown by Menezes at al.~\cite{Menezes}  
such an interaction can be reduced to the phonon-mediated 
local hybridization that can also lead to the formation 
of Bose-Einstein condensate under the supposition that 
it will be sufficiently strong, what depends on values
of electron-phonon constants. 

\subsection{Mean-field results}
Since not all of the above mentioned candidates are the one-dimensional 
or quasi one-dimensional systems, it is interesting to ask if the above 
obtained results persist also in higher dimensions. Unfortunately, 
the DMRG method is not a very convenient tool for the study of lattice systems 
in $D>1$. Therefore, it is  necessary to choose another method that 
should be able to describe sufficiently precise the excitonic 
effects in $d$-$f$ systems. From this point of view a very promising 
method seems to be the HF approximation with the 
CDW instability. Indeed, we have showed in our
previous paper~\cite{Fark2}, that this method is able to describe almost 
perfectly the ground-state phase diagram of the extended Falicov-Kimball 
model in $D=2$, including the existence of excitonic phase.       

Let us briefly summarize the basic steps of this approximation.
In the presence of CDW instability, the order parameters can be 
written as follows~\cite{Brydon, Fark2}:
\begin{eqnarray}
\langle n_i^f\rangle=n^f+\delta_f\cos({\bf Q}\cdot {\bf r}_i)\ ,\\
\langle n_i^d\rangle=n^d+\delta_d\cos({\bf Q}\cdot {\bf r}_i)\ ,\\
\langle f_i^+d_i\rangle=\Delta+\Delta_P\cos({\bf Q}\cdot {\bf r}_i)\ ,
\end{eqnarray}
where $\delta_{d}$ and $\delta_{f}$ are the order parameters of the CDW state
for the $d$- and $f$-electrons, $\Delta$ is the excitonic average and
${\bf{Q}}=\pi$ (${\bf{Q}}=(\pi,\pi)$) is the nesting vector for 
$D=1$ ($D=2$).

Using these expressions the HF Hamiltonian of the generalized
Falicov-Kimball model ($H=H_0+H_{t_f}+H_V+H_{non}$) is
\begin{eqnarray}
{\cal H}&=&-t_d\sum_{\langle i,j\rangle} d^+_id_j
           -t_f\sum_{\langle i,j\rangle}f^+_if_j
           + E_f\sum_i n_i^f
           + U\sum_i (n_f+\delta_f\cos({\bf Q}\cdot {\bf r}_i))n_i^d \\
          &+&U\sum_i (n_d+\delta_d\cos({\bf Q}\cdot {\bf r}_i))n_i^f
           + \sum_{ij} (V_{ij} -U[\Delta+\Delta_P\cos({\bf Q}\cdot {\bf
r}_i)]\delta_{ij})d_i^+f_j+H.c. \nonumber
\end{eqnarray}

This Hamiltonian can be  diagonalized by the following
canonical transformation
\begin{eqnarray}
\begin{array}{ccc}
\gamma_k^m=u_k^md_k + v_k^md_{k+{\bf Q}} + a_k^mf_k + b_k^mf_{k+{\bf Q}}\ ,
&   &
m=1,2,3,4\ ,
\end{array}
\end{eqnarray}
where $\Psi_k^m=(a_k^m,b_k^m,u_k^m,v_k^m)^T$ are solutions of the associated
Bogoliubov-de Gennes eigenequations:

\begin{equation}
H_k\Psi_k^m=E_k^m\Psi_k^m\ ,
\end{equation}
with
\begin{eqnarray}
H_k=\left( \begin{array}{cccc}
    \epsilon_k^d+Un_f &  U\delta_f &   V_k-U\Delta &     -U\Delta_P \\
     U\delta_f  & \epsilon_{k+Q}^d+Un_f & -U\Delta_P &    V_{k+Q}-U\Delta\\
     V^*_k-U\Delta^*   &  -U\Delta^*_P   &  \epsilon_{k}^f+Un_d+E_f & U\delta_d \\
     -U\Delta^*_P &  V^*_{k+Q}-U\Delta^*    &  U\delta_d  & \epsilon_{k+Q}^f+Un_d+E_f\\
            \end{array}
\right),
\end{eqnarray}
and the  corresponding dispersions $\epsilon_k^d$, $\epsilon_k^f$ and $V_k$ 
are obtained by the Fourier transform of the $d$/$f$-electron
hopping amplitudes and the local/nonlocal hybridization. 
Then the HF parameters $n_d,\delta_d,n_f,\delta_f,\Delta,\Delta_P$ and 
the density of zero-momentum excitons $n_0$ can be expressed directly in terms 
of the  Bogoliubov-de Gennes eigenvectors
$a_k^m,b_k^m,u_k^m,v_k^m$~\cite{Brydon}.

To verify the ability of this method to describe the formation of the 
Bose-Einstein condensate we have calculated firstly the density
of the zero momentum excitons as functions of model parameters $V,t_f$
and $E_f$ for the one dimensional case. The results of our numerical
calculations are displayed in Fig~4 (the left panels) and they show 
that the HF approximation with the CDW instability is able
to decribe all aspects of formation of zero-momentum condensate 
in $D=1$ discussed above within the DMRG method. 
Although, there are some small qualitative as well as quantitative 
differences, like the discontinuous changes of $n_0$ with $t_f$ 
near $t_f=0$, or almost two times larger values of $V_c$ in comparison
to DMRG studies (see the inset in the a) panel), in principle, 
the DMRG and HF pictures are same.
In accordance with DMRG results we have found a strong enhancement 
of $n_0$ near $E_f \sim -1.5$ also in the HF solutions. 
Performing the numerical derivative of $n_d$ with respect to $E_f$ 
and comparing it with the behaviour of $n_0(E_f)$ one can easily verify 
(see the instet in the e) panel)  that this enhancement is indeed 
connected with changes in the occupation of $f$ ($d$) orbitals
as conjectured above on the base of $DMRG$ results.
\begin{figure}[h!]
\begin{center}
\includegraphics[width=7.5cm]{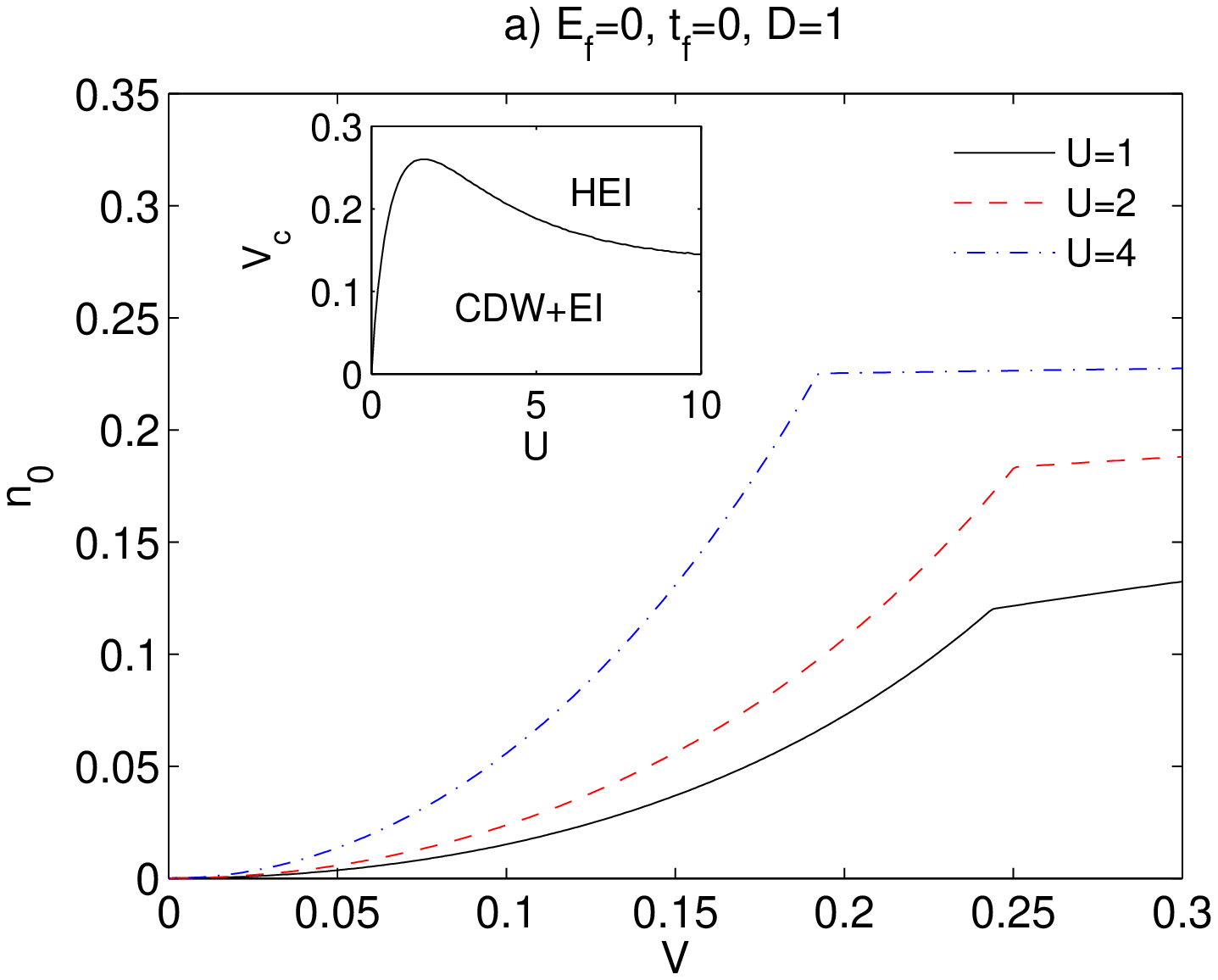} \hspace{-0.4cm}
\includegraphics[width=7.5cm]{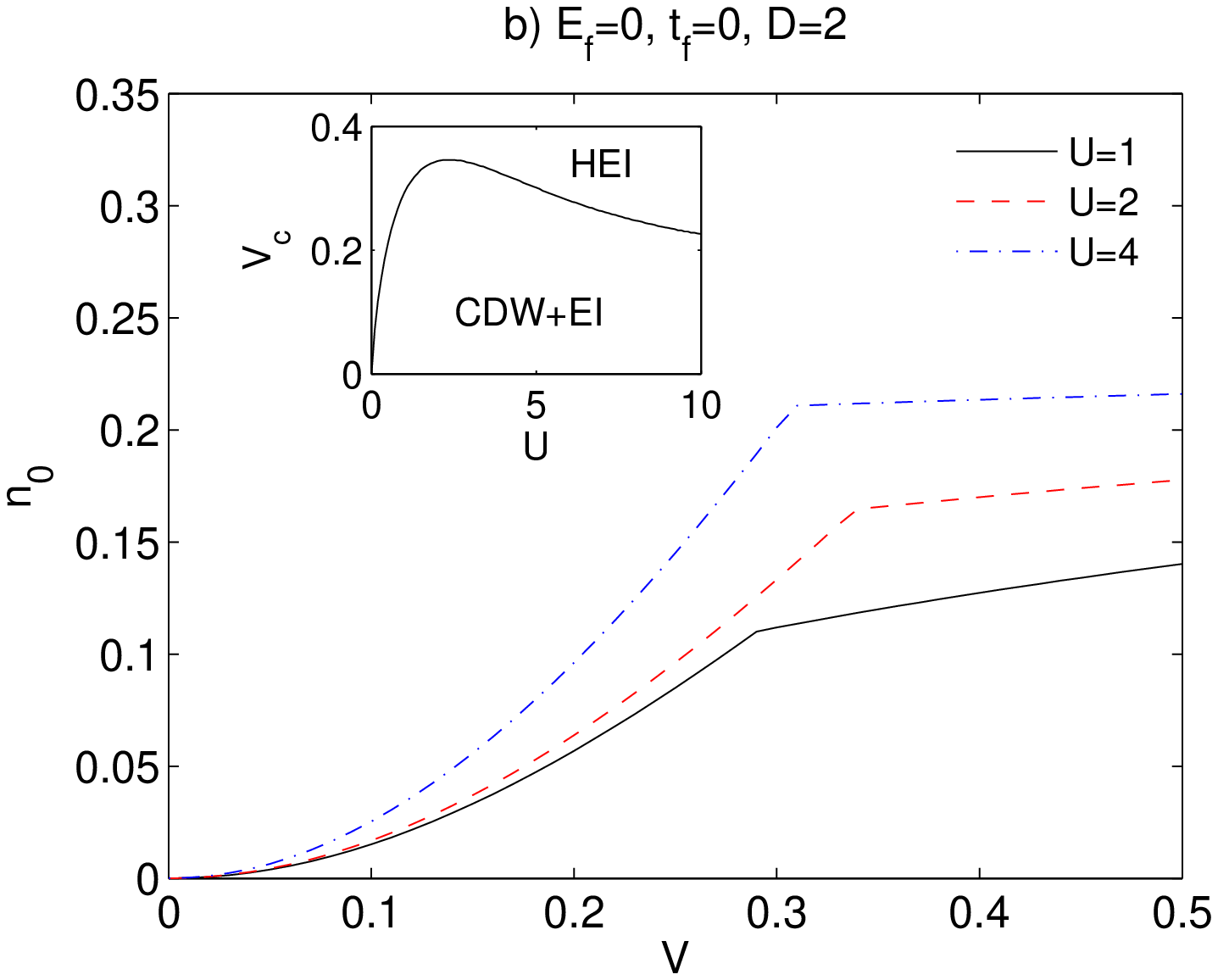}
\includegraphics[width=7.5cm]{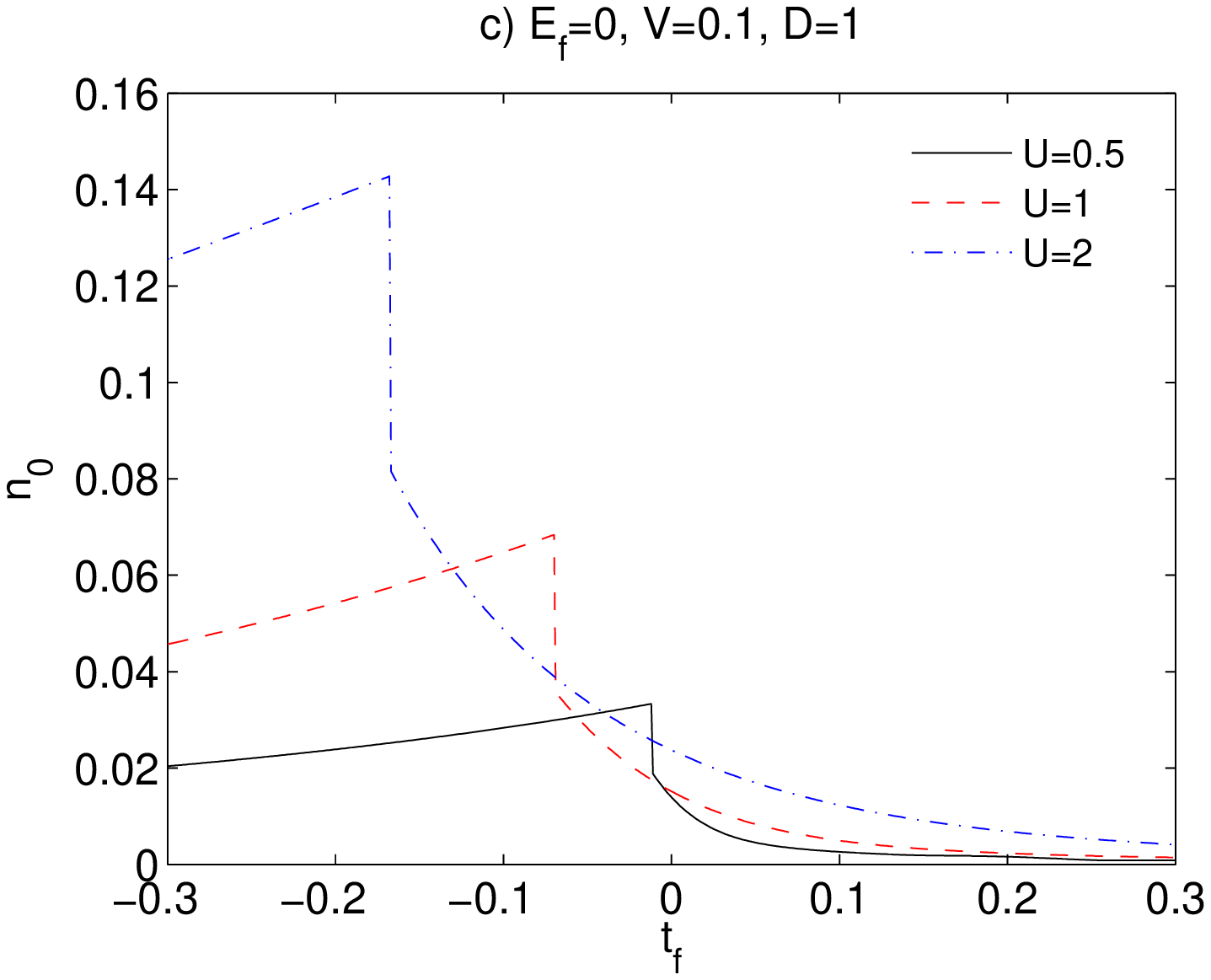} \hspace{-0.4cm}
\includegraphics[width=7.5cm]{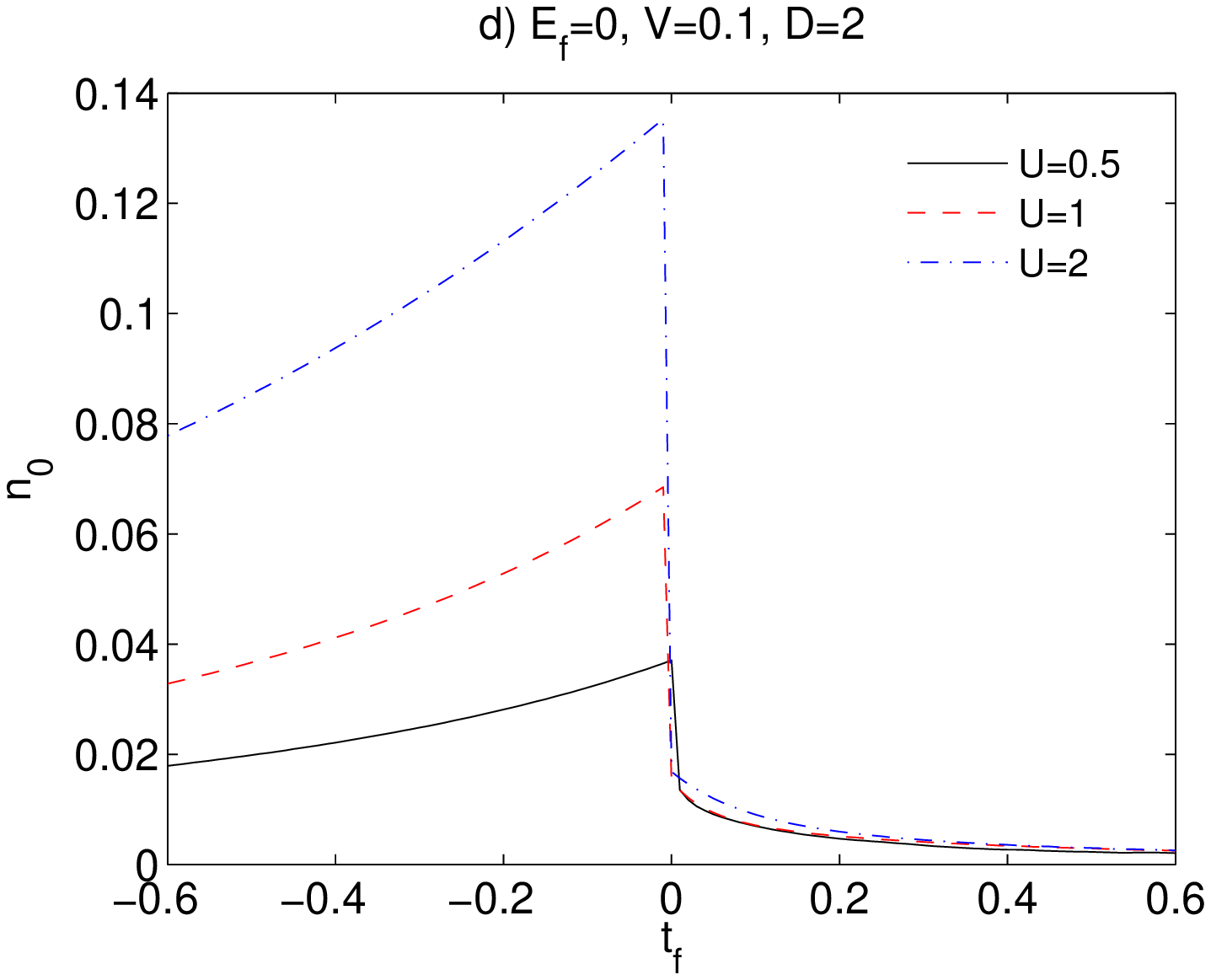}
\includegraphics[width=7.5cm]{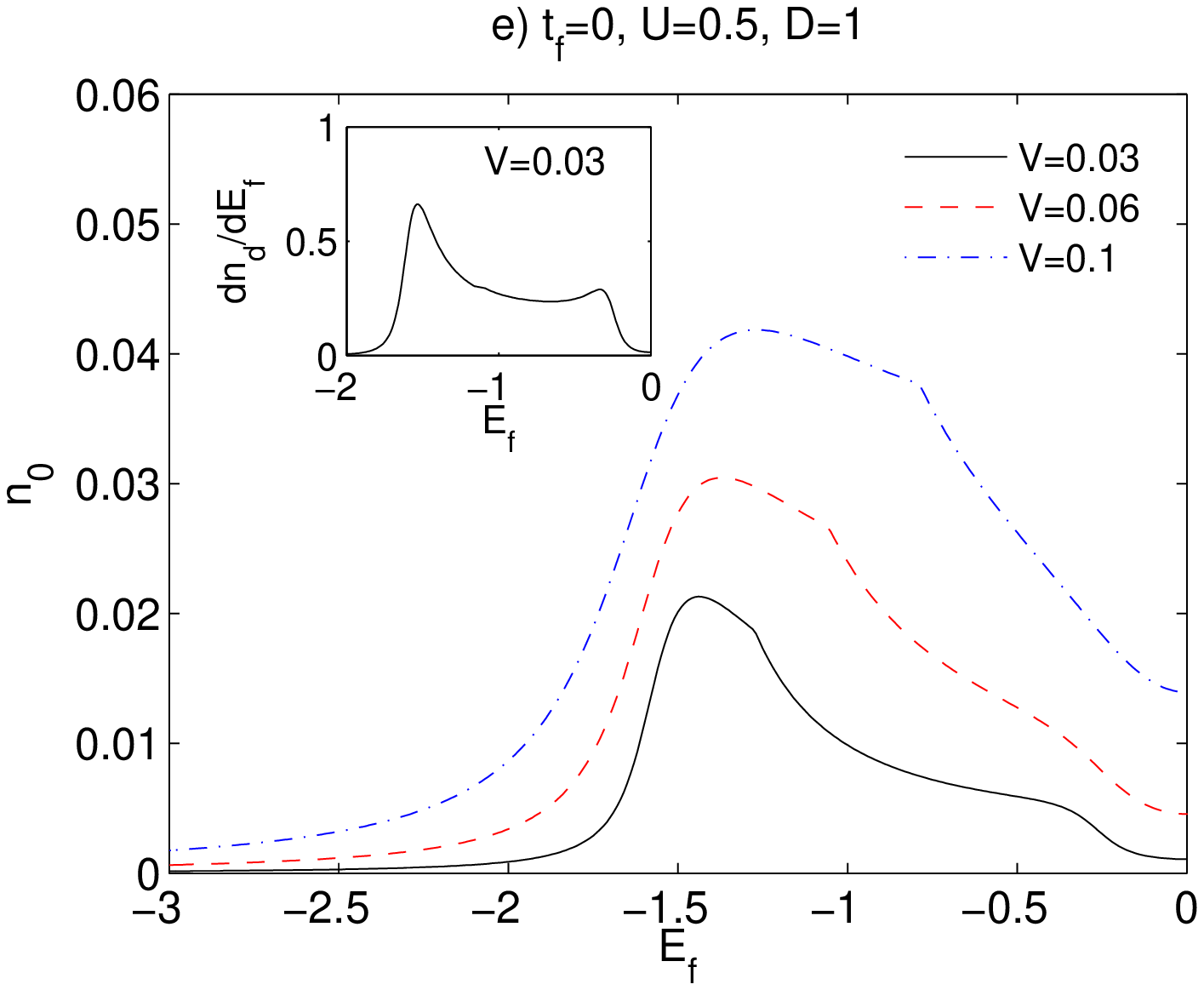} \hspace{-0.4cm}
\includegraphics[width=7.5cm]{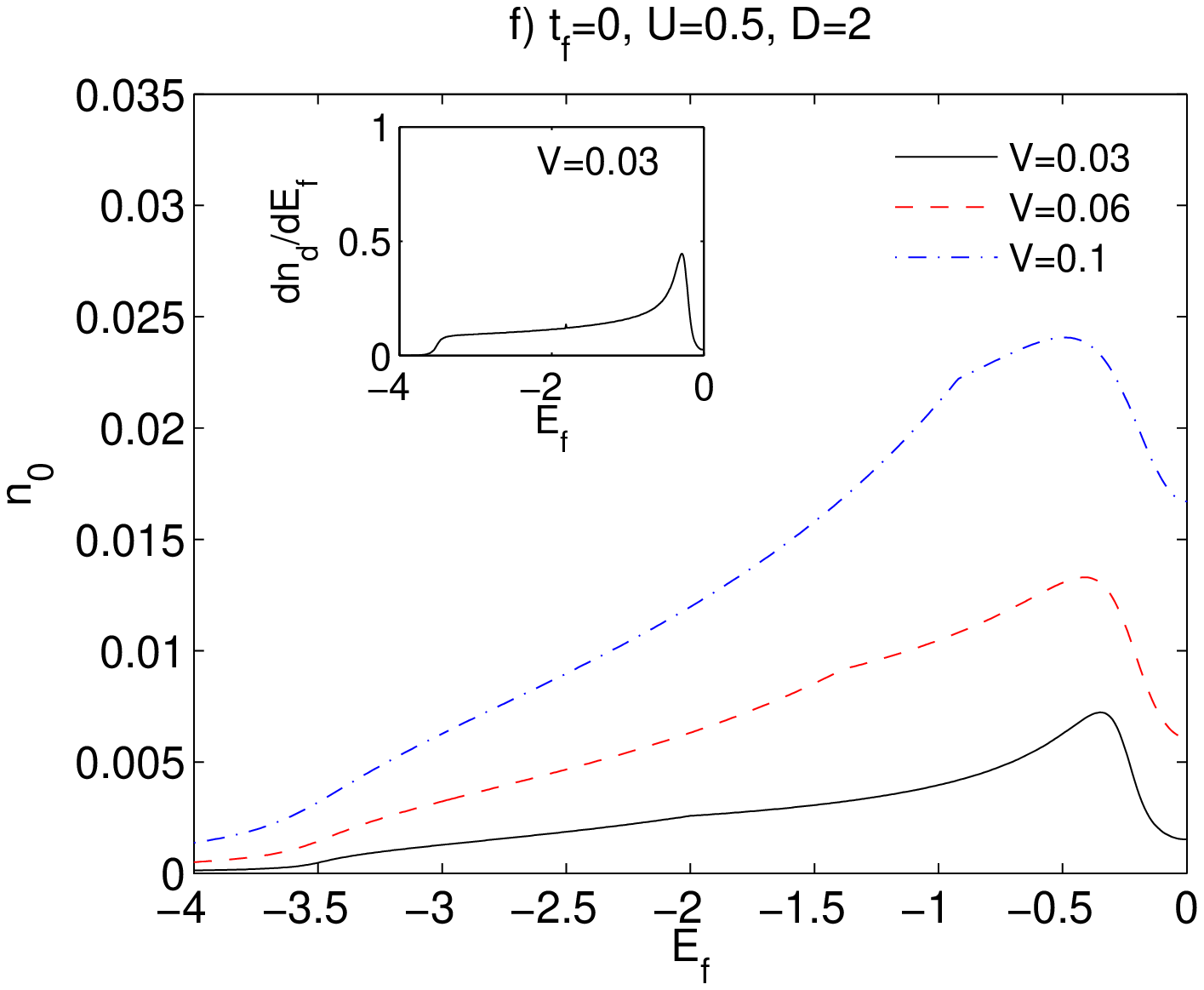}
\end{center}
\vspace*{-0.8cm}
\caption{\small $n_0$ as a function of $V,t_f, E_f$ calculated by HF
approximation with the $CDW$ instability in the one (the left panels)
and two dimensions (the right panels). The insets in the panels a) and b)
show the phase boundary between the homogeneous excitonic insulator (HEI) phase 
and the coexisting excitonic insulator and CDW state (CDW+EI).}
\label{fig4}
\end{figure}

The same calculations we have performed also in D=2. 
The resultant behaviours of the density of zero momentum excitons 
$n_0$ as functions of $V,t_f$ and $E_f$ are plotted in Fig.~4 
(the right panels) 
and they clearly confirm that with increasing dimension of the system 
the zero-momentum condensate remains robust. In comparison to the 
one-dimensional results there are two differences, and namely,
that the discontinuous changes of $n_0$ (as a function of $t_f$) take place 
now strictly at $t_f=0$ in all examined cases and that the maximum
in $n_0(E_f)$ shifts to higher values of $E_f$, where the periodic
solutions with CDW instability minimize the ground state energy. 
And finally, it should be noted that in accordance with DMRG results 
we have found no sign of formation of Bose-Einstein condensate 
for the case of nonlocal hybridization with the inversion symmetry 
neither in our HF solutions for both $D=1$ as well as  $D=2$.

In conclusion, we have examined effects of various factors, like  
$f$-electron hopping, the local and nonlocal hybridization, as well as 
the increasing dimension of the system on the formation and condensation 
of excitonic bound states in the generalized Falicov-Kimball model.
It was found that the negative values of the $f$-electron hopping
integrals $t_f$ support the formation of zero-momentum condensate,
while the positive values of $t_f$ have the fully opposite effect. The 
opposite effects on the formation of  condensate exhibit
also the local and nonlocal (with the inversion symmetry) hybridization.
The first one strongly supports the formation of zero-momentum condensate, 
while the second one destroys it completely. 
Moreover, it was found that in the pressure induced case 
($E_f \sim p$), the model is able to describe,
at least qualitatively, the increase in the total density of 
excitons $n_T$ with external pressure and the increase or deccrease
(according to the initial position of $E_f$ at ambient pressure)
in $n_0$ and $n^{un}_d$. And finaly, it was shown (by HF studies) 
that the zero-momentum condensate remains robust with increasing dimension 
of the system.

\vspace{0.5cm}
This work was supported by Slovak Research and Development Agency (APVV)
under Grant APVV-0097-12 and ERDF EU Grants under the contract No.
ITMS 26220120005 and ITMS26210120002.

\end{document}